\documentclass[twocolumn,english,aps,prb,twocolum,superscriptaddress,bibnotes,amsmath,amssymb,floatfix]{revtex4-1}
\usepackage[colorlinks=true,citecolor=blue,linkcolor=blue]{hyperref}
\usepackage{xcolor}
\usepackage{soul}
\usepackage[utf8]{inputenc}
\usepackage[english]{babel}
\usepackage{amsmath,amsfonts,amssymb}
\usepackage[T1]{fontenc}
\usepackage{url}
\usepackage{amsmath}
\usepackage{amsfonts}
\usepackage{amssymb}
\usepackage{graphicx}

\begin{document}
\title{OH absorption in on-chip high-Q resonators}

\author{Lue Wu}
\affiliation{T. J. Watson Laboratory of Applied Physics, California Institute of Technology, Pasadena, California 91125, USA}

\author{Maodong Gao}
\affiliation{T. J. Watson Laboratory of Applied Physics, California Institute of Technology, Pasadena, California 91125, USA}

\author{Jin-Yu Liu}
\affiliation{T. J. Watson Laboratory of Applied Physics, California Institute of Technology, Pasadena, California 91125, USA}

\author{Hao-Jing Chen}
\affiliation{T. J. Watson Laboratory of Applied Physics, California Institute of Technology, Pasadena, California 91125, USA}

\author{Kellan Colburn}
\affiliation{T. J. Watson Laboratory of Applied Physics, California Institute of Technology, Pasadena, California 91125, USA}

\author{Henry A. Blauvelt}
\affiliation{T. J. Watson Laboratory of Applied Physics, California Institute of Technology, Pasadena, California 91125, USA}

\author{Kerry J. Vahala}
\email[]{vahala@caltech.edu}
\affiliation{T. J. Watson Laboratory of Applied Physics, California Institute of Technology, Pasadena, California 91125, USA}

\maketitle

\noindent\textbf{Abstract}

\noindent  {\bf Thermal silica is a common dielectric used in all silicon-photonic circuits. And bound hydroxyl ions (Si-OH) can provide a significant component of optical loss in this material on account of the wet nature of the thermal oxidation process. A convenient way to quantify this loss relative to other mechanisms is through OH-absorption at 1380 nm. Here, using ultra-high-Q thermal-silica wedge microresonators, the OH absorption loss peak is measured and distinguished from the scattering loss base line over a wavelength range from 680 nm to 1550 nm. Record-high on-chip resonator Q factors are observed for near-visible and visible wavelengths, and the absorption limited Q factor is as high as 8 billion in the telecom band. OH ion content level around 2.4 ppm (weight) is inferred from both Q measurements and by Secondary Ion Mass Spectroscopy (SIMS) depth profiling. }

\medskip

High-Q integrated resonators have become an essential component in nonlinear photonics. Most often, the guided light in these structures has a significant fractional overlap with silica (e.g., all-silica wedge resonators \cite{lee2012,LeoBQ_OL20} and silica-clad ultra-low-loss silicon nitride waveguides \cite{jin2021hertz,puckett2021422}). It is therefore important to understand the loss limits imposed by the silica used in silicon photonic processing. Besides interface scattering loss, optical absorption from bound hydroxyl ions (Si-OH) can be a significant component of loss \cite{MDabsQ_NC22}, especially since thermal silica is prepared using a process involving water. Bound hydroxyl ions produce a well-known fundamental absorption peak at 2720 nm \cite{HUMBACH199619} and the overtone at 1380 nm is used here to measure OH absorption loss in ultra-high-Q thermal-silica wedge microresonators. Further comparison to scattering loss is made over a wavelength range from 680 nm to 1550 nm. Also, using cavity-enhanced photo-thermal spectroscopy \cite{MDabsQ_NC22} near the 1380 nm band, the OH ion content level is estimated to be 2.4 ppm (weight). This value also agrees with Secondary Ion Mass Spectroscopy (SIMS) depth profiling performed on the resonator material. Outside of the 1380 nm band, scattering loss is confirmed by measurement and modeling to be the dominant loss mechanism in the samples tested.

\begin{figure*}[t!]
\centering
\includegraphics[width = \linewidth]{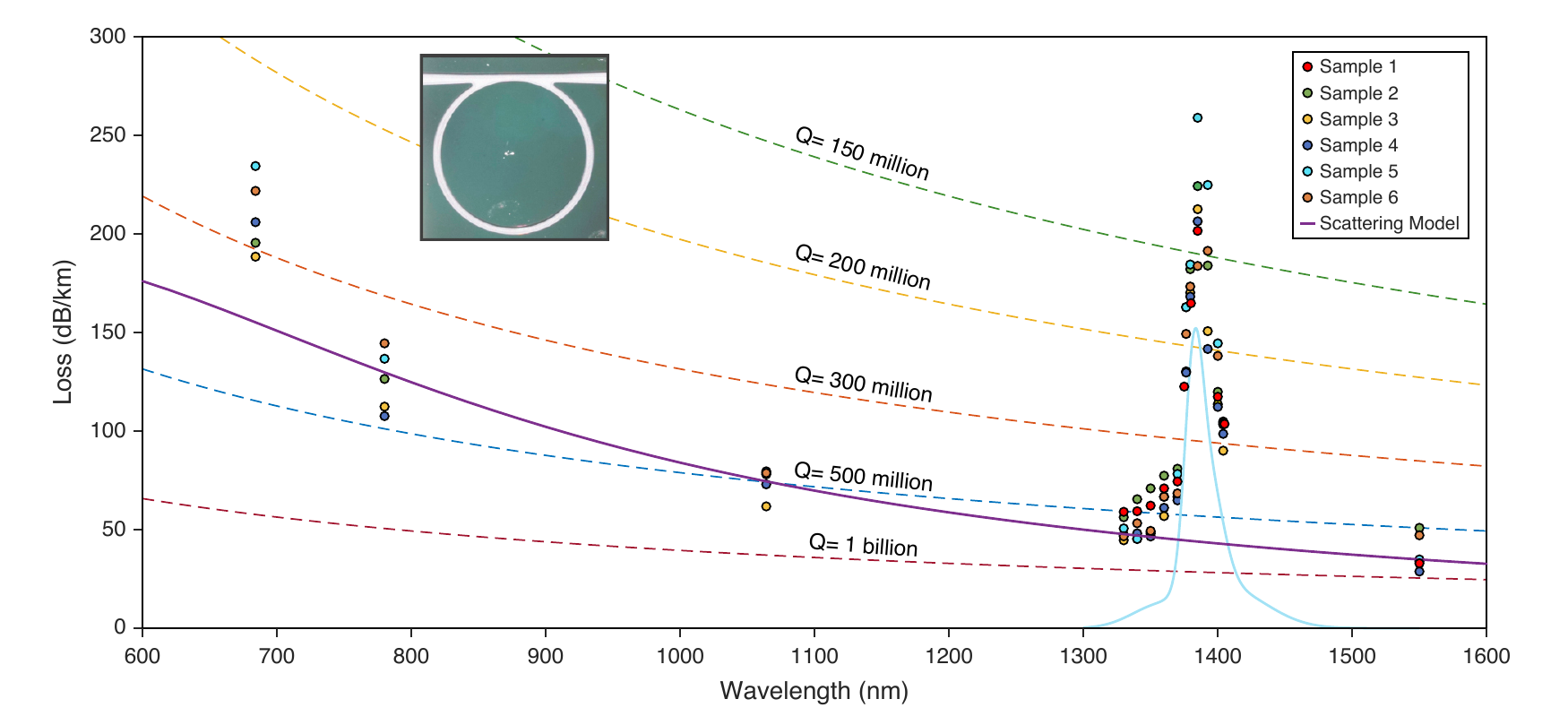}
 \caption{ Optical loss measured from 680 nm to 1550 nm reveals the existence of the OH absorption peak near 1380 nm.  The data are obtained by optical Q-factor measurement. The purple solid line is the predicted loss from surface-roughness scattering (see text). The light blue solid line is the estimated OH absorption loss in 1380 nm band (corresponding to 2.4 ppm weight). Dashed contours correspond to different Q-factor values.
Inset: typical optical micrograph of the optical resonator used in the measurements.}
\label{Fig1}
\end{figure*}

Thermally-grown silica wedge whispering-gallery resonator devices were prepared as measurement samples, and featured 8 $\mu$m thermal oxide thickness with resonator diameter 6.5 mm (10 GHz free spectral range, FSR). The fabrication steps are detailed in reference \cite{LeoBQ_OL20} and a photo micrograph of a typical resonator is the Fig. \ref{Fig1} inset. 
The thermal silica was grown from float-zone (low background doping level) silicon wafers using the wet oxidation method. As a final step, the devices were annealed for 18-hours at 1000$^{\circ}$C in N$_{2}$. The samples were stored and measured in a dry N$_2$ environment to minimize environmental impact on optical loss. Tapered fiber couplers \cite{Cai2000,spillane2003ideality} were used to couple probe light to the resonator samples.

Fig. \ref{Fig1} summarizes the optical loss measurements using six of these devices measured at wavelengths from 680 nm to 1550 nm. All resonators feature intrinsic quality factor (Q-factor) over 500 million at 1550 nm. The corresponding Q-factors are also given in the plot (dashed contours). Significantly, the devices have record-high Q values for on-chip devices in the shorter wavelength bands: 300 million at 685 nm, 500 million at 780 nm, and 600 million at 1064 nm. As an aside, the  loss per unit length $\alpha$ (dB/km) is determined from the optical Q-factor measurement using the relationship $  \alpha=4343\times \frac{2\pi n}{\lambda Q_0}\; \textrm{(dB/km)}$, where Q$_0$ is intrinsic Q-factor in billions, $n$ is the mode effective refractive index and $\lambda$ is the wavelength (nm).
The intrinsic Q-factor Q$_0$ is obtained from the relationship $1/Q_0= 1/Q_L - 1/Q_c$ by measuring the resonance linewidth with Lorentzian lineshape fitting to obtain the loaded Q-factor Q$_L$ followed by measurement of transmission depth to infer the coupling Q-factor Q$_c$.

A strong increase in loss near 1380 nm is apparent in all six samples, corresponding to the OH absorption band and reaching over 200 dB/km. This absorption quickly decreases for wavelengths above and below 1380 nm. The loss in other spectral regions is believed to be dominated by Rayleigh scattering. The wavelength dependence of this scattering within the resonator mode volume would scale approximately as $\lambda^{-4}$, and does not fit the data. Modeling of surface scattering is described in the Methods and provides better agreement with the wavelength dependence. This theoretical dependence is given by the purple curve where surface roughness variance (1.9 nm) and correlation length (350 nm) are assumed in the plot \cite{Barwicz:05}.  The origin of the increased loss at the shortest wavelengths measured is not known, but possibilities include absorption loss from metallic ion impurities and scattering from material density fluctuations. For example, SIMS \cite{RWilsionSIMS1989}  data has shown that Chromium is a residual contaminant in our processing, and Cr$^{3+}$ ions can contribute 1.6 dB/km/ppbw at 800 nm (peak at 625 nm) to absorption \cite{RevModPhys.51.341}.  

\begin{figure}[t!]
\centering
\includegraphics[width = \linewidth]{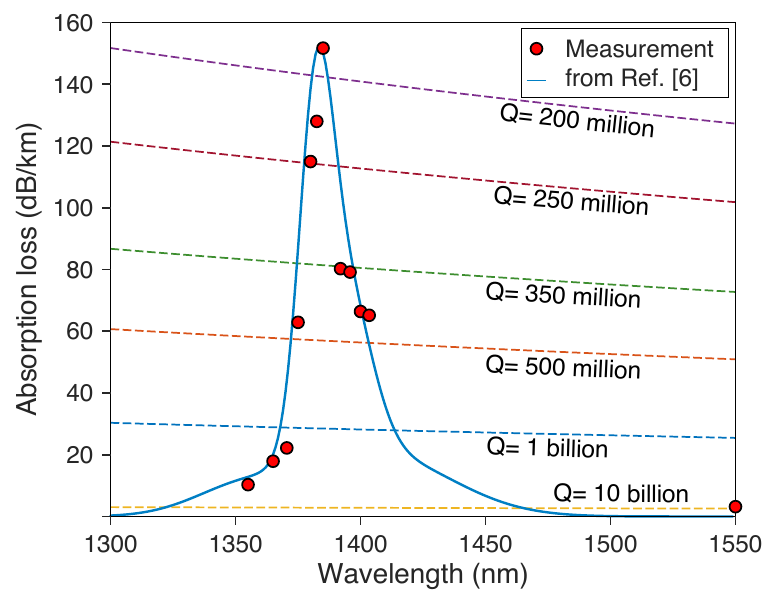}
\caption{ {\bf  Absorption loss measurement results. } Absorption loss as measured by cavity-enhanced photothermal spectroscopy in the 1380 nm band. Red dots: data from Sample 1 in  Fig. \ref{Fig1}. Blue solid line: 2.4 ppm (weight) OH content level absorption lineshape based on reference \cite{HUMBACH199619}.}  
\label{Fig2}
\end{figure} 

Cavity-enhanced photothermal spectroscopy (CEPS) \cite{MDabsQ_NC22} was used to further study the OH absorption loss. This method measures the microresonator resonance thermal triangle formation induced by the thermo-optical effect \cite{carmon2004dynamical} to determine mode volume absorption. Details on this method are provided in reference \cite{MDabsQ_NC22}. Fig.  \ref{Fig2} summarizes the wavelength dependence of the measured absorption loss in both the 1380 nm band and at 1550 nm.  The measured absorption near 1380 nm follows reasonably well the OH overtone lineshape in silica as determined elsewhere \cite{HUMBACH199619} (blue curve superimposed in plot). To fit the magnitude of the lineshape function to the data, an OH content level of 2.4 ppm weight is used. Also, the data near 1550 nm indicate that by further reduction in scattering loss, the existing thermal silica can provide absorption limited Q factors as high as 8 billion. This value is 2$\times$ larger than measured for wedge resonators in a previous study \cite{MDabsQ_NC22} and is attributed to application of an improved resist cleaning step. The absorption loss measurement results at 1385 nm and 1550 nm are summarized in Table \ref{table:1} and compared with values from the literature. 

\begin{table}[h!]
\centering
\begin{tabular}{ |c|c|c|} 
\hline
Material  &  1385 (dB/km) &  1550 (dB/km)\\\hline
thermal silica in this study & 152 & 3.2 \\\hline
1 ppmw OH in silica \cite{HUMBACH199619}  & 62.7  &  \\\hline
Wet fiber  studied in \cite{HUMBACH199619} & 48500  & 100  \\\hline
Fiber studied in \cite{thomas2000towards} & 1.172  & 0.045  \\\hline
low-OH fiber \cite{li1985optical,keiser2010fiber}  & 0.05  &  \\\hline
\end{tabular} 
\caption{Summary of \textbf{absorption loss rates} at both 1385 nm and 1550 nm from Fig.\ref{Fig2} and taken from the literature.. Note that ppmw is parts per million in weight (equivalently, $\mu$g/g) } 
\label{table:1}
\end{table}

\begin{figure}[h!]
\centering
\includegraphics[width = \linewidth]{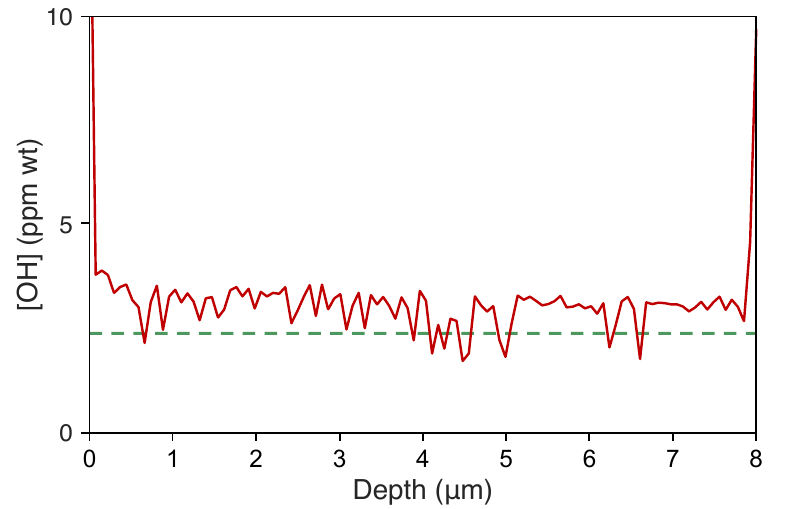}
\caption{OH ion depth distribution measured using secondary ion mass spectroscopy (SIMS) over the resonator thickness. Dashed line indicates 2.4 ppmw level inferred from data in Fig. \ref{Fig2}.}  
\label{Fig3}
\end{figure}

To further confirm the OH ion impurity level and also study its spatial distribution over the resonator oxide thickness, secondary ion mass spectroscopy (SIMS) depth profiling \cite{RWilsionSIMS1989} was employed (see Fig. \ref{Fig3}). The absolute OH ion concentration in this measurement is obtained from the ratio $^{17}$OH/$^{30}$Si, as calibrated against reference samples  \cite{HAURI200299,10.2138/am.2011.3810}. The OH concentration remains nearly constant along the depth direction with a value in reasonable agreement with the CEPS results in Fig. \ref{Fig2}. This uniformity is believed to result from diffusion-driven equilibrium \cite{thomas2000towards} during the long oxide growth process and the post high-temperature annealing process \cite{LeoBQ_OL20}. The higher OH concentration near the surfaces is consistent with the hydrophilic nature of silica causing water to be adsorbed  on the surfaces \cite{gorodetsky1996ultimate,rokhsari2004loss}. 

In summary, using thermal silica wedge microresonators and by measuring optical loss  from 680 nm to 1550 nm, the OH ion absorption peak in 1380 nm band stands out from  base line scattering loss. Absorption-only loss in the 1380 nm band was also measured using cavity enhanced photo-thermal spectroscopy. This gave a maximum absorption loss of 152 dB/km at 1385 nm. Fitting of the absorption spectrum gives an OH content level of 2.4 ppm weight, which is consistent with separate measurements performed on the resonator oxide by Secondary Ion Mass Spectroscopy. 
Outside the OH overtone absorption band, surface scattering was determined to be the dominant loss mechanism. Record high Q factors (limited by scattering) were measured for visible and near-visible wavelengths. Also, with further reduction scattering loss, Q values approaching 8 billion are feasible at 1550 nm. This value is higher than determined in previous measurements \cite{MDabsQ_NC22}. Overall, this investigation highlights the potential for even lower loss levels and higher Q factors using thermal silica in photonic devices.

\section*{Methods}

The scattering loss for silica wedge resonators is dominant by surface inhomogeneity scattering rather than bulk inhomogeneity scattering, as evident by the following two experimental observations. First, the optical loss is roughly inversely proportional to the bending radius \cite{LeoBQ_OL20}, a characteristic feature for surface roughness scattering \cite{gorodetsky1996ultimate,Gorodetsky:00}. Second, with bending radius over 1 mm, the resonances of wedge resonators never have splitting effect \cite{KippenbergsplittingOL} even with Q-factor over 1 billion \cite{LeoBQ_OL20}. 

Surface scattering loss can be modeled as dipole radiation of surface inhomogeneities polarized by the local electrical field \cite{Barwicz:05}. The equivalent current density induced by the corresponding dipole moments is
$$\vec J(\vec r)=-i\omega \epsilon_0(n_0^2-n_1^2) \vec E(\vec r) $$
where $\omega$ is the angular frequency of the light, $\epsilon_0$ is the free space permittivity, $n_0,n_1$ are refractive indices of cavity and cladding materials and $\vec E(\vec r)$ is the local electrical field. The corresponding radiation field and Poynting vector are given by

$$\vec{A}(\vec{r})=\frac{\mu e^{in_1 k_0 r}}{4\pi r}\oint e^{-in_1 k_0\hat{r}\cdot\vec{r} }J(\vec{r'})h(\vec{r'})dS'$$
$$\mathcal{\vec S}(\vec r)=\hat r\frac{\omega n_1k_0}{2\mu}|\hat r\times \vec A(\vec{r})|^2$$
where $k_0$ is the wave number of the light and $h(\vec{r})dS$ is the volume element of inhomogeneity. When adding up the radiation field from different part of surface, their coherence is limited by the correlation length. Using the correlation function of surface roughness $\langle h(x)h(y)\rangle=\sigma^2 R(\frac{x-y}{B})$, and assuming that the correlation length $B$ is small enough to ignore the variation of the field within this length, the Poynting vector field can be reduced to 
$$\mathcal{\vec S}(\vec r)=\hat r\frac{\mu \omega n_1k_0}{2(4\pi r)^2}\sigma^2B^2\oint |\vec{J}(\vec{r'})\times \hat r|^2\tilde{R}[\vec\beta(\vec{r'})-n_1k_0\hat r_\|(\vec{r'})]dS'$$
where $\tilde R(\vec k)$ is Fourier transformation of $R(\vec x)$, $\vec\beta(\vec{r'})$ is the propagation constant of cavity mode at $\vec{r'}$, and $\hat r_\|(\vec{r'})$ is the component of $\hat r$ parallel to the surface at $\vec{r'}$.

In the calculation of $\mathcal{\vec S}$ and the radiation power, the parameters of the cavity mode, such as mode profiles and propagation constants, are acquired by numerical simulation (Lumerical MODE). The correlation function of surface roughness in the calculation is assumed to be a Gaussian function with the form $R(\frac{x-y}{B})=\frac{1}{B\sqrt{2\pi}}e^{-\frac{(x-y)^2}{2B^2}}$. The parameters of the surface roughness are acquired by Atomic Force Microscopy (AFM) measurements, fitted with Gaussian-type function, revealing that the lower surface has  $\sigma$ of 2 nm, whereas the upper and wedge surfaces have $\sigma$ smaller than 0.5 nm \cite{lee2012,lee2012ultra}.

\noindent
\medskip
\begin{footnotesize}

 \noindent \textbf{Funding Information}: 

 \noindent
This work was funded by DARPA (HR0011-22-2-0009).

\vspace{3mm}

\noindent \textbf{Acknowledgments}: 

\noindent
The authors acknowledge support provided by The Kavli Nanoscience Institute (KNI) at Caltech and Caltech Microanalysis Center (CMC) at Division of Geological $\&$ Planetary Sciences. The authors thank George R. Rossman, Yunbin Guan and Qing-Xin Ji for fruitful discussions. 

\medskip

\noindent \textbf{Disclosures:} The authors declare no conflicts of interest.

\end{footnotesize}

\bibliography{paper}






\end{document}